\documentclass[sigconf]{acmart}
\usepackage{subcaption}

\AtBeginDocument{%
  }

\setcopyright{acmlicensed}
\copyrightyear{2018}
\acmYear{2018}
\acmDOI{XXXXXXX.XXXXXXX}

\acmConference[NIP@IR 2025]{New Interaction Paradigms for Information Retrieval in the Era of Generative AI}
{July 17, 2025}
{Padua, Italy}

\acmISBN{978-1-4503-XXXX-X/2018/06}

\begin{document}


\title{Personality over Precision: Exploring the Influence of Human-Likeness on ChatGPT Use for Search}



\author{Mert Yazan}
\authornotemark[1]
\email{m.yazan@hva.nl}
\orcid{0009-0004-3866-597X}
\affiliation{%
  \institution{Amsterdam University of Applied Sciences}
  \city{Amsterdam}
  \country{Netherlands}
}

\author{Frederik Bungaran Ishak Situmeang}
\email{f.b.i.situmeang@uva.nl}
\orcid{0000-0002-2156-2083}
\affiliation{%
  \institution{Amsterdam University of Applied Sciences}
  \city{Amsterdam}
  \country{Netherlands}
}

\author{Suzan Verberne}
\email{s.verberne@liacs.leidenuniv.nl}
\orcid{0000-0002-9609-9505}
\affiliation{%
  \institution{University of Leiden}
  \city{Leiden}
  \country{Netherlands}
}


\renewcommand{\shortauthors}{Yazan et al.}

\begin{abstract}

    Conversational search interfaces, like ChatGPT, offer an interactive, personalized, and engaging user experience compared to traditional search. On the downside, they are prone to cause overtrust issues where users rely on their responses even when they are incorrect. What aspects of the conversational interaction paradigm drive people to adopt it, and how it creates personalized experiences that lead to overtrust, is not clear.
    To understand the factors influencing the adoption of conversational interfaces, we conducted a survey with 173 participants. 
    We examined user perceptions regarding trust, human-likeness (anthropomorphism), and design preferences between ChatGPT and Google. To better understand the overtrust phenomenon, we asked users about their willingness to trade off factuality for constructs like ease of use or human-likeness. 
    Our analysis identified two distinct user groups: those who use both ChatGPT and Google daily (DUB), and those who primarily rely on Google (DUG). The DUB group exhibited higher trust in ChatGPT, perceiving it as more human-like, and expressed greater willingness to trade factual accuracy for enhanced personalization and conversational flow. Conversely, the DUG group showed lower trust toward ChatGPT but still appreciated aspects like ad-free experiences and responsive interactions. Demographic analysis further revealed nuanced patterns, with middle-aged adults using ChatGPT less frequently yet trusting it more, suggesting potential vulnerability to misinformation. 
    Our findings contribute to understanding user segmentation, emphasizing the critical roles of personalization and human-likeness in conversational IR systems, and reveal important implications regarding users' willingness to compromise factual accuracy for more engaging interactions.
 
\end{abstract}




\maketitle


\section{Introduction}



    There has been a growing interest in making information retrieval (IR) more conversational \cite{conv_info_seeking}. Conversational search offers significant advantages over traditional search by enabling context-aware, natural language interactions that support personalized responses and enhanced user engagement \cite{conv_info_seeking}. Conversational search interfaces, such as ChatGPT, are rated as more enjoyable and useful compared to traditional search engines \cite{ux_chatgpt_google, winds_of_change}. The ever-increasing influence of Large Language Models (LLMs) allows conversational search to be even more powerful, as they can be used in many tasks like query reformulation, search clarification, conversational retrieval, and response generation \cite{survey_conversationalsearch}. 

    Despite ChatGPT's popularity, Google has approximately 14 to 16 billion daily searches \cite{google_daily}, whereas ChatGPT processes around 1 billion prompts each day \cite{chatgpt_stats}. Given the fact that ChatGPT or similar platforms are commonly used for tasks like coding and writing besides search \cite{arena_explorer, claude_tasks, why_use_chatgpt}, traditional search engines are still the most popular means of information finding. Yet, the current paradigm is steering towards a more unified experience. Apps like Perplexity or the search feature of ChatGPT provide LLMs with real-time web search results to mitigate inaccuracies. Also, Google has integrated generative AI into its search \cite{google_genai_search}, which shows a clear trend of making search more conversational. 

    Perceived usefulness is one of the main pillars of technology acceptance \cite{TAM}, and anthropomorphism (human-likeness) is shown to increase acceptance in customer-support bots \cite{rethink_conv_styles}. Similarly, in the case of ChatGPT, users highlight human-like qualities \cite{trust_chatgpt_google} and perceived usefulness \cite{ux_chatgpt_google, winds_of_change} as key reasons for choosing it. Users who interact with ChatGPT report higher enjoyment and satisfaction, compared to users who interact with Google, despite similar task performances \cite{ux_chatgpt_google}. Apparently, the conversational interaction paradigm driven by the human-likeness and perceived usefulness leads to more enjoyable experiences \cite{ux_chatgpt_google, psychological_distance}. Similarly, trust in ChatGPT is influenced positively by conversationality, fast responses, and perceived human-likeness \cite{trust_chatgpt_google}.

    Despite the advantages, there is an emerging issue with conversational interfaces: overtrust. While the generative nature of LLMs empowers them as effective conversational tools, it also makes them prone to hallucinations and fabricating inaccurate information \cite{hallucination}. Despite inconsistencies and a worse performance than Google in fact-checking tasks, people who only used ChatGPT were more likely to trust its answers \cite{ux_chatgpt_google}. Users show overreliance on conversational AI interfaces, even when it is incorrect \cite{conv_info_seeking}. Anthropomorphism reduces risk perception \cite{rethink_conv_styles}, which can further aggravate overtrust. Therefore, even though users trust chat-based interfaces for their interaction flow and personality, they become more susceptible to misinformation. Vulnerable groups like older adults are at a higher risk, since they are more prone to believe fake information \cite{fakenews_aging}. 

    To better understand the adoption of conversational interfaces, we conduct an exploratory survey study with 173 participants to identify the types of users who interact with Google and ChatGPT. We investigate the main design principles that drive tool preferences and how they relate to perceived human-likeness and trust. We further look into how design, trust, and human-likeness relate to people's perception of factuality, while identifying people's willingness to trade off factuality for ease-of-use and more personalized interactions. Finally, an age‑ and gender‑based analysis further illuminates how demographic factors modulate the aforementioned factors.
    
    Our findings reveal two user groups: people who use both ChatGPT and Google daily, and people who only use Google daily. The groups diverged significantly in their trust towards the platforms, as daily users of both platforms showed more trust towards ChatGPT and perceived it as more human-like. Both groups listed personalized outputs as the most common reason that they would prefer ChatGPT over Google. Interaction flow and conversational interface emerged as the most popular design aspects of ChatGPT, especially for daily users of both platforms. When asked if they would agree on trading factuality for ease-of-use and human-likeness, the groups again diverged, as daily users of both agreed more with the statement. Age groups also suggest an interesting pattern: middle-aged (30-55) adults use ChatGPT less frequently but trust it more. Therefore, the main contributions of this paper are:
\begin{enumerate}
    \item We provide distinct user segments (based on usage patterns and age) that have significantly different perceptions and preferences towards ChatGPT. 
    \item We identify the influence of human-likeness and personalization on trust and design preferences in conversational assistants.
    \item We demonstrate that users may be willing to trade off factuality for more personalized interactions.
\end{enumerate}

\section{Background}

    Various studies identify a higher technology acceptance of conversational search compared to traditional search, due to its usefulness and ease-of-use \cite{winds_of_change, psychological_distance, ux_chatgpt_google}. Both those dimensions are pillars of the technology acceptance model (TAM) \cite{TAM}, and a higher acceptance leads to a higher use intention and adoption \cite{winds_of_change}. The user interface is directly related to technology acceptance, as design choices affect perceived ease-of-use and usefulness \cite{TAM}. Conversational search interfaces offer an interactive and personalized experience over traditional methods, reducing the cognitive and emotional load of the users, leading to increased acceptance \cite{emotional_load}. Compared to web search, users preferred conversational assistants more when the task felt psychologically distant (far in time/space/likelihood/social closeness) \cite{psychological_distance}. Low effort, natural, and interactive dialogues were listed as the main reasons behind the user's preferences \cite{psychological_distance}. Both \citet{psychological_distance} and \citet{emotional_load}'s work verifies the link between design choices and technology acceptance, and sheds light on why people prefer conversational interfaces over traditional search. 

    Anthropomorphism (human-likeness) is one of the dimensions of interface design. Chatbots can be designed to be more human-like, by manipulating their visual appearance (e.g., with avatars), or their communication style (emoji usage, emotional responses) \cite{anthro_cues}. Human-device interactions with conversational voice assistants mirror human-human relationships \cite{va_anthro}, and anthropomorphism increases technology acceptance of chatbots \cite{rethink_conv_styles}. For conversational search, human-likeness not only increases acceptance, but it also affects trust positively \cite{trust_chatgpt_google}. People start to form relationships with conversational assistants due to anthropomorphism, and their trust levels increase \cite{va_anthro}. 

    The downside of trust and usefulness provided by conversational search interfaces is the emerging issue of overtrust \cite{ux_chatgpt_google, conv_xai}. People who only interact with ChatGPT report higher trust despite inconsistent answers, compared to people who only interacted with Google \cite{ux_chatgpt_google}. Conversational explainable AI interfaces lead to higher user engagement and trust, but also greater over-reliance on AI, even when it is incorrect \cite{conv_xai}. Despite its benefits, anthropomorphism might play an indirect role in the overtrust issue. People list perceived human-likeness as one of the factors that increase their trust in ChatGPT \cite{trust_chatgpt_google}. High anthropomorphism increases forgiveness of chatbot errors \cite{chatbot_forgiveness}, and risk perception of customers decreases when they interact with a more human-like system \cite{rethink_conv_styles}. Given the fact that people use ChatGPT for sensitive topics like medical information \cite{why_use_chatgpt}, this makes the overtrust issue an even bigger problem. Especially groups that are more prone to believing misinformation, like older adults, are at a higher risk \cite{fakenews_aging}. People with high openness to technology perceive ChatGPT to be more useful than others \cite{winds_of_change}, which might put them at a bigger risk of misinformation.

    Despite its widespread adoption, the reasons why people prefer ChatGPT over search engines, like Google, are not well-established. Specifically, the role design and anthropomorphism play in acceptance, and how they may lead to overtrust issues, are not clear. \citet{ux_chatgpt_google} identified that people overtrust ChatGPT, but they did not explore the causes or potential implications. Previous literature shows a positive relationship between trust, human-likeness, and acceptance, but how those relationships lead to negative consequences like overtrust is unknown. Furthermore, it is not clear how aware users are about the overtrust issue. Some user segments, like younger adults with high digital literacy, might be more aware of the pitfalls of generative AI, whereas older adults might not question the information they receive because they don't have the same digital literacy levels. To bridge this gap, in this study, we explore the overtrust issue, its potential causes (human-likeness), and how it affects different user segments.
    




   \begin{figure}[ht]
        \centering
        \includegraphics[width=\linewidth]{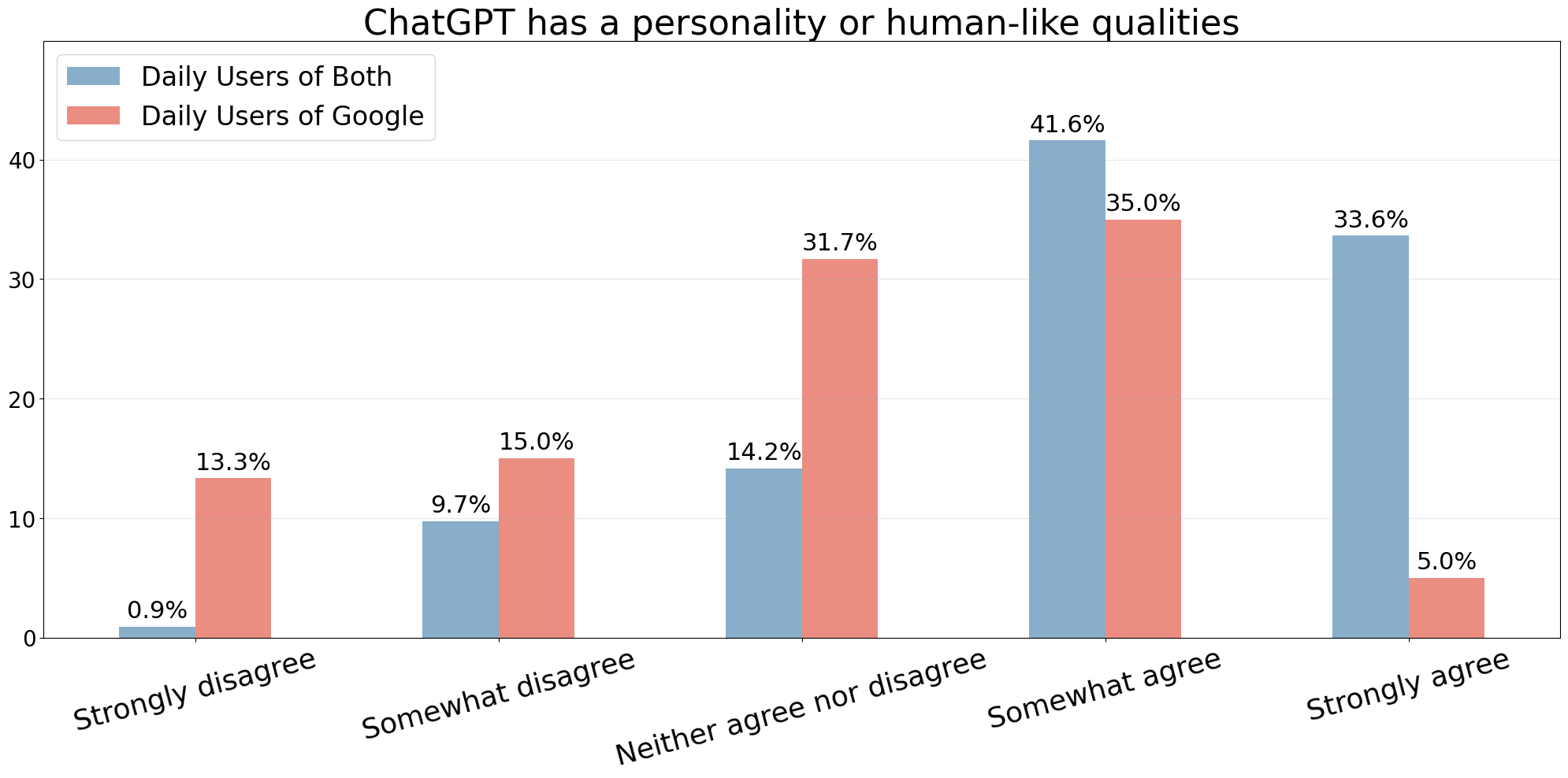}
        \caption{Distribution of people agreeing that ChatGPT has human-like qualities. DUB has higher agreement rates, but DUG also agrees more with the statement rather than disagreeing.}
        \label{fig:personality}
        \Description{Bar chart showing the distribution of agreement levels with the statement that ChatGPT has human-like qualities, split between DUB and DUG groups. The DUB group shows a higher percentage of participants agreeing strongly or somewhat. The DUG group also shows more agreement than disagreement overall, but with a more balanced distribution across the response scale.}
    \end{figure}

\section{Methods}

    To understand how design factors and human-likeness are associated with trust and acceptance, potentially leading to overtrust, we conducted a survey with 200 participants. We compared the two main interaction paradigms in web search: a) traditional search engines (Google), and b) conversational, chatbot-based tools (ChatGPT). We used Qualtrics for survey design and Prolific for participant recruitment. To control for cultural differences, we restricted our participant pool to the US and applied a gender quota. Using Prolific's custom filters, we chose only people who have previous experience with ChatGPT and Google. 

    Upon entering the survey, participants were informed that the survey is about information search, and not about other use-cases of ChatGPT (creative writing, coding, etc). Even though we specifically mentioned Google and ChatGPT, we informed participants that they can answer questions if they use similar alternatives (e.g., Perplexity instead of ChatGPT, Bing instead of Google). Participants are asked how frequently (daily, weekly, monthly, never) they use both tools. Even with the pre-filtering, we noticed that 8 people claimed to never use ChatGPT. Therefore, we did not include them in the final analysis.

    Usage frequencies showed the emergence of two main groups: Daily Users of Google (DUG) and Daily Users of Both (DUB). DUB group uses both platforms actively, while DUG only uses Google actively while using ChatGPT on a weekly or monthly basis. There is a small group of people who claim to use ChatGPT daily, but not Google. Since they were a minority of 9 people, they did not provide enough samples to draw a conclusion from. Therefore, they were excluded from the final analysis. 
    
    After usage frequency questions, participants moved on to assess trust and human-likeness, with a 5-point Likert scale of agreement. Participants are asked how much they agree that each platform is trustworthy/human-like. We also added a binary choice question about which platform they trust more to see which one they ultimately prefer. With this study, we aim for an exploratory approach rather than developing predictive models. Given this exploratory focus and the need to minimize survey fatigue, we opted for single-item measures to assess trust and human-likeness. Prior work suggests that single-item measures can be sufficient—and even preferable—when comparing well-known alternatives, as they minimize survey fatigue and capture participants’ immediate, overall impressions \cite{single_item}. 

    In the next block, participants answered design-related questions. They were given multiple-choice questions about what design aspects of ChatGPT they like compared to Google, and vice versa. We constructed the options using the previous findings from the literature. For example, people identified transparent sources as an advantage for Google \cite{psychological_distance}, while natural, low-effort interactions are chosen for ChatGPT \cite{psychological_distance, trust_chatgpt_google}. Finally, participants were asked how much they would agree to prefer ease-of-use, human-likeness, or conversationality gained by ChatGPT despite the risk of incorrect or misleading information. The full set of survey questions can be found in Appendix~\ref{survey_questions}.

    We further filtered out participants who had null answers, even though the questions were mandatory. Also, 5 of the participants didn't have either gender or age information, and they were removed, too. After filtering, we had 173 participants in total. Based on the age distribution of our participants, we grouped them into four brackets: 18–30, 30–40, 40–55, and 55+. These brackets were chosen to ensure a balanced representation across groups while also reflecting meaningful life stages that may influence technology use and trust perception. 

    \begin{table}[t]
        \centering
        \small
        \resizebox{\columnwidth}{!}{%
        \begin{tabular}{l|cccc|c}
        \toprule
        & \textbf{18--30} & \textbf{30--40} & \textbf{40--55} & \textbf{55+} & \textbf{Total} \\
        \hline
        \textbf{Male} & & & & & \\
        \hspace{1em}DUB & 9 (16.7\%) & 20 (37.0\%) & 19 (35.2\%) & 6 (11.1\%) & 54 \\
        \hspace{1em}DUG & 4 (13.3\%) & 10 (33.3\%) & 9 (30.0\%) & 7 (23.3\%) & 30 \\
        \hspace{1em}Overall & 13 (15.5\%) & 30 (35.7\%) & 28 (33.3\%) & 13 (15.5\%) & \textbf{84} \\
        \hline
        \textbf{Female} & & & & & \\
        \hspace{1em}DUB & 22 (37.3\%) & 15 (25.4\%) & 12 (20.3\%) & 10 (16.9\%) & 59 \\
        \hspace{1em}DUG & 5 (16.7\%) & 6 (20.0\%) & 10 (33.3\%) & 9 (30.0\%) & 30 \\
        \hspace{1em}Overall & 27 (30.3\%) & 21 (23.6\%) & 22 (24.7\%) & 19 (21.3\%) & \textbf{89} \\
        \hline
        \textbf{All} & & & & & \\
        \hspace{1em}DUB & 31 (27.4\%) & 35 (31.0\%) & 31 (27.4\%) & 16 (14.2\%) & 113 \\
        \hspace{1em}DUG & 9 (15.0\%) & 16 (26.7\%) & 19 (31.7\%) & 16 (26.7\%) & 60 \\
        \hspace{1em}Overall & \textbf{40} (23.1\%) & \textbf{51} (29.5\%) & \textbf{50} (28.9\%) & \textbf{32} (18.5\%) & \textbf{173} \\
        \hline
        \bottomrule
        \end{tabular}
        } 
        \caption{Participant distribution by gender and age across user groups, with row-wise percentages. DUB refers to Daily Users of Both, and DUG refers to Daily Users of Google.}
        \label{tab:demographics}
    \end{table}

    A breakdown of demographics by age, gender, and daily user groups can be found in Table~\ref{tab:demographics}. Looking at the demographics, we see that 65\% of the participants fall into the DUB group. 51.5\% of the final participants are female, and there is a similar gender distribution between groups; ~66\% females and ~64\% males are in DUB. Age groups show a differing pattern, as 55+ participants have a higher representation in the DUG. While the DUB participant distribution is similar between 18-55 age groups, it falls sharply for 55+. Contrarily, the 18-30 group is less represented in DUG. 

\section{Results}

    \begin{table*}[]
        \centering
        \begin{tabular}{lccc}
        \toprule
        \textbf{Metric} & \textbf{Mean (DUB)} & \textbf{Mean (DUG)} & \textbf{$p$‑value$^{\dagger}$} \\
        \midrule
        I find Google to be trustworthy. & 4.168 & 4.000 & 0.150\\
        I find ChatGPT to be trustworthy. & 4.301 & 3.467 & $<\!0.001$*\\
        ChatGPT has a personality or human-like qualities. & 3.974 & 3.033 & $<\!0.001$*\\
        I would prefer ChatGPT over Google because it is easier to use, \\
        \hspace{2cm} despite incorrect or misleading information.   & 3.540 & 2.383 & $<\!0.001$*\\
        I would prefer ChatGPT over Google because it is humanlike, \\
        \hspace{2cm} despite incorrect or misleading information.    & 3.522 & 2.717 & $<\!0.001$*\\
        I would prefer ChatGPT over Google because it is conversational, \\
        \hspace{2cm} despite incorrect or misleading information.    & 3.478 & 2.683 & $<\!0.001$*\\
        \bottomrule
        \multicolumn{4}{l}{\footnotesize $^{\dagger}$Mann–Whitney $U$ test (two‑tailed). “*” marks results significant at $\alpha=.05$.}\\
        \end{tabular}
        \caption{Inter‑group comparison between “Daily Users of Both” (DUB) and “Daily Users of Google” (DUG)}
        \label{tab:trust_and_personality}
    \end{table*}

   \begin{figure}[]
        \centering
        \begin{subfigure}{\linewidth}
            \centering
            \includegraphics[width=\linewidth]{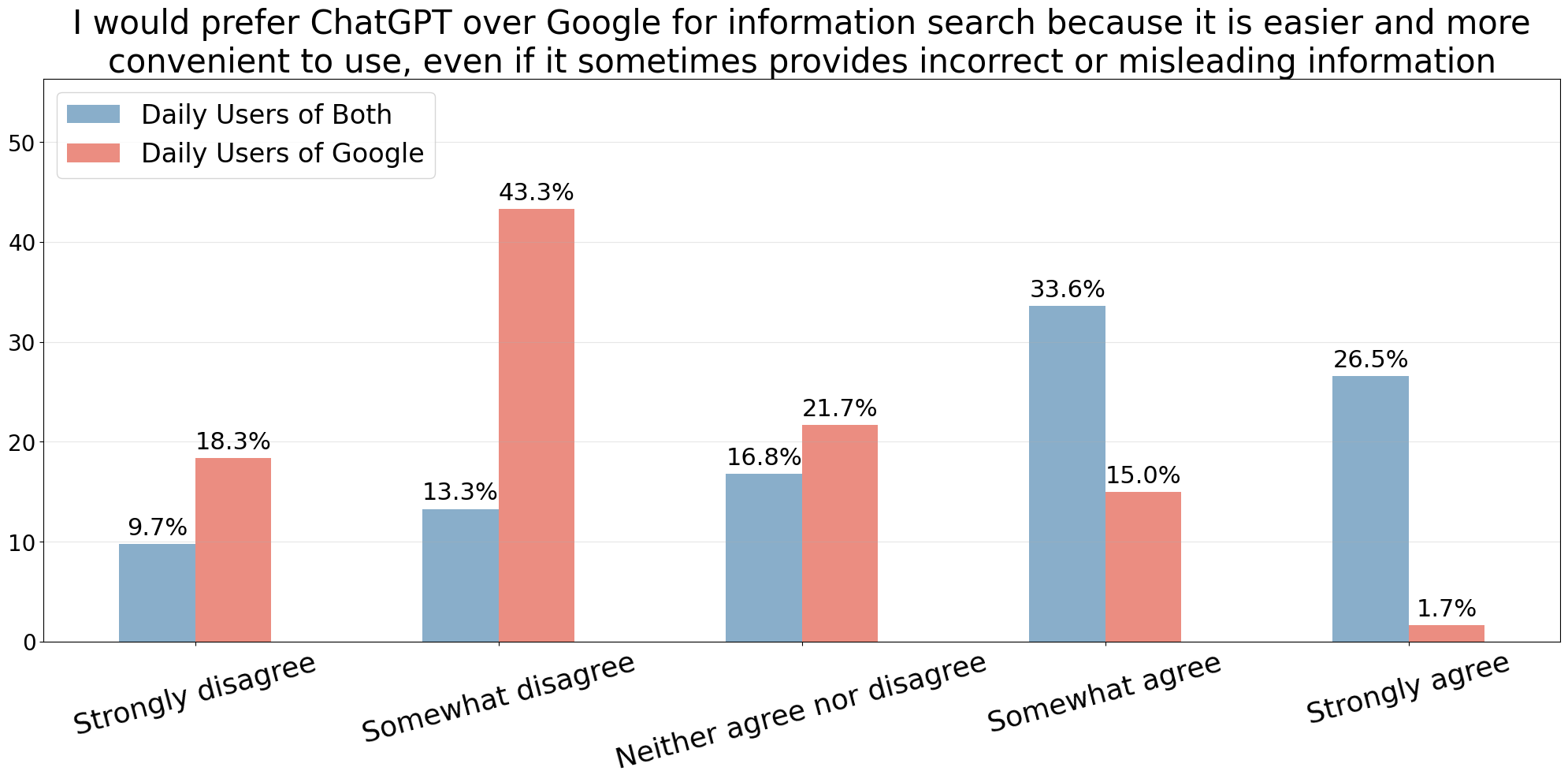}
            \caption{Preference for ease of use over factuality}
            \label{fig:tradeoff_easeofuse}
        \end{subfigure}
    
        \begin{subfigure}{\linewidth}
            \centering
            \includegraphics[width=\linewidth]{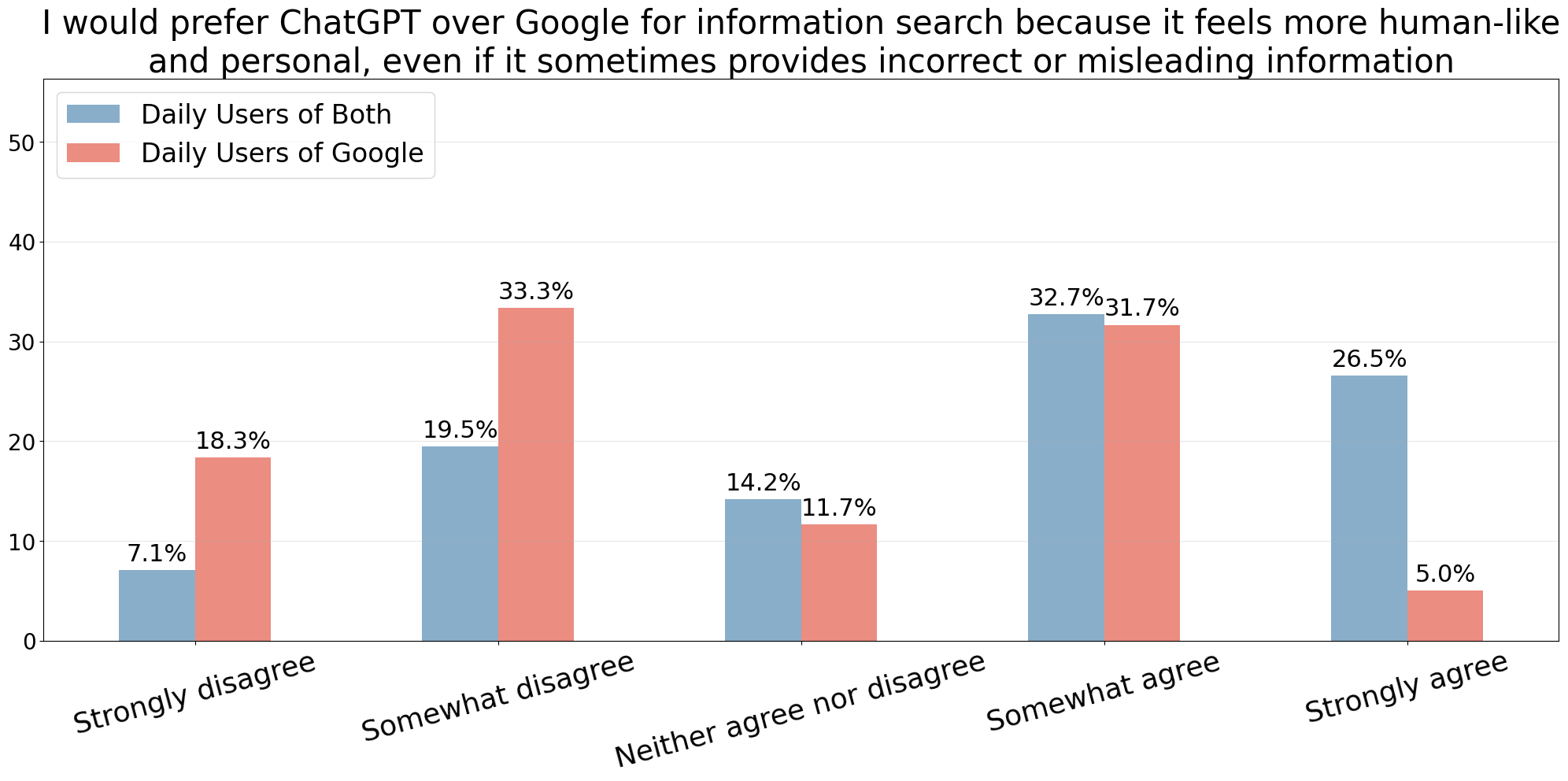}
            \caption{Preference for human-likeness over factuality}
            \label{fig:tradeoff_humanlikeness}
        \end{subfigure}
    
        \begin{subfigure}{\linewidth}
            \centering
            \includegraphics[width=\linewidth]{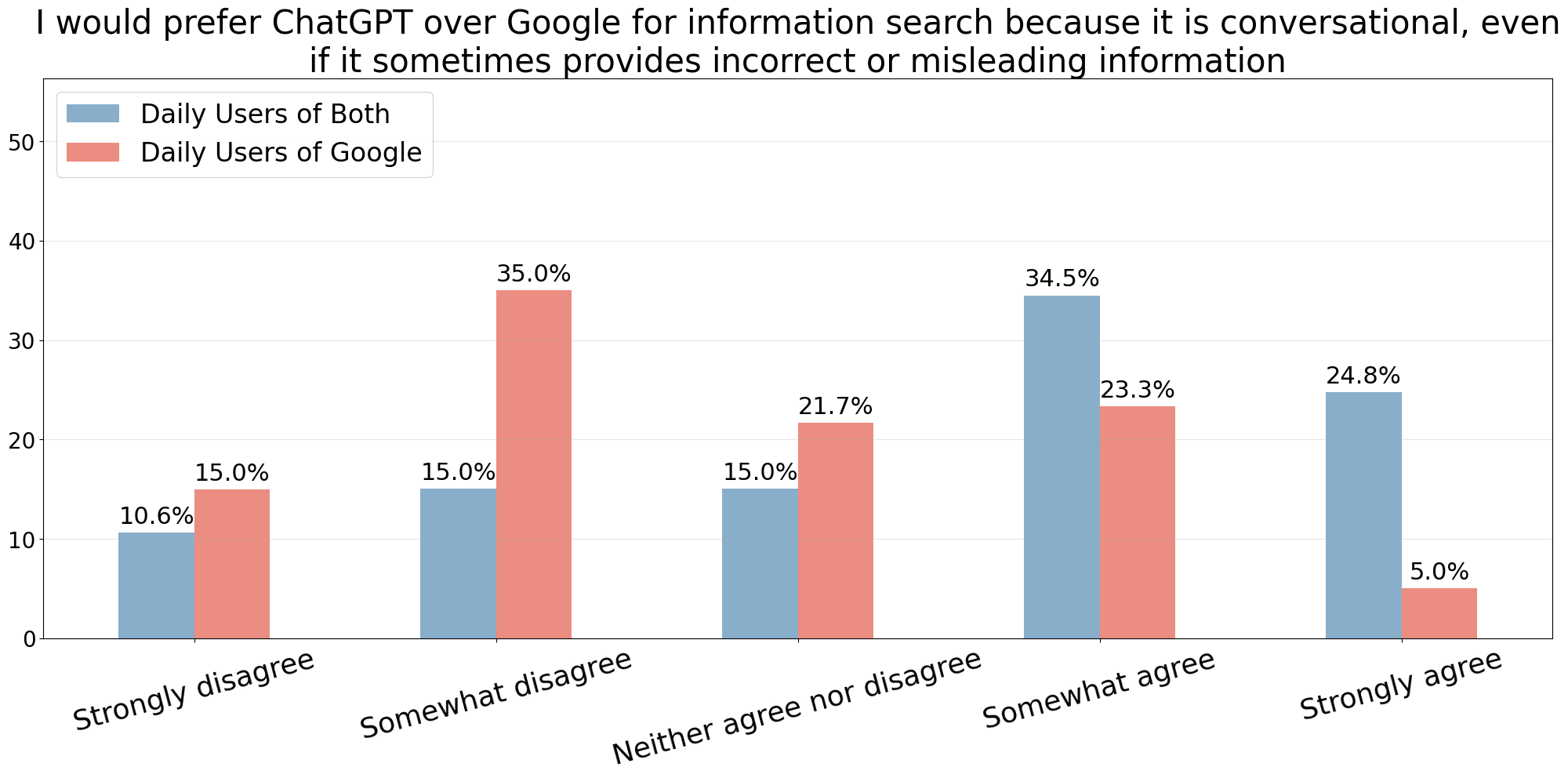}
            \caption{Preference for conversationality over factuality}
            \label{fig:tradeoff_conversationality}
        \end{subfigure}
    
        \caption{Comparison between user groups for trade-off questions. DUB preferences don't differ significantly, while the DUG group agrees more with trading of factuality with human-likeness and conversationality, instead of ease-of-use.}
        \label{fig:trade_off}
        \Description{Three bar charts comparing two user groups—Daily Users of Both (DUB) and Daily Users of Google (DUG)—on their willingness to trade factual accuracy for different ChatGPT features. The first chart shows trade-off for ease-of-use, the second for human-likeness, and the third for conversationality. DUB scores remain consistently high across all, while DUG shows relatively higher agreement for human-likeness and conversationality than for ease-of-use.}
    \end{figure}
    
    \subsection{Trust and Human-likeness}
    
    To check for any significant differences between the groups, we used the Mann-Whitney U test because it is robust for uneven group sizes and requires no assumption of normality. First, we analyzed how trust, human-likeness, and factuality trade-offs differ between user groups. Table~\ref{tab:trust_and_personality} shows no significant difference between the groups in their trust towards Google, but a significant difference emerges for ChatGPT. DUG users showed lower levels of trust for the platform, and overwhelmingly answered that they trust Google more (83\%), while DUB favored ChatGPT (58\%). Interestingly, DUB not only trusts ChatGPT more but also Google (DUB mean: 4.17 vs. DUG mean: 4.0). DUB perceived ChatGPT to be human-like more than DUG users, as a significant difference emerged between their answers. Yet, according to Figure~\ref{fig:personality}, even 40\% of the DUG group agreed that ChatGPT is human-like, compared to the 28\% that disagreed. 

    Table~\ref{tab:trust_and_personality} shows a significant divergence between groups for trade-off questions. Following their lower trust levels, daily users of Google are less likely to trade off factuality. The DUB group is more willing to trade off factuality for ease-of-use, conversationality, and for more personalized interactions provided by ChatGPT, and their agreement levels between the 3 constructs don't change significantly. Interestingly, DUG agrees more on trading factuality for human-like and conversational interactions, compared to ease-of-use. Table~\ref{tab:DUG_pairs} shows a significant difference between trade-off items for DUG when the Wilcoxon signed‑rank test is applied. Figure~\ref{fig:trade_off} highlights the differing attitudes between groups and questions. 

    \subsection{Design Preferences}

    Personalization and human-likeness emerge as a theme again when participants are asked about the reasons they would prefer ChatGPT over Google; both groups chose personalized responses as the main reason. DUB listed clearer answers as the second most important factor, while speed for certain types of queries was more important for DUG. Out of the 63 participants of DUG, only 9 of them chose "None", indicating that there are some aspects of ChatGPT they like despite their preference for Google. The differences between groups were higher for items related to ease-of-use and human-likeness. 7 people chose ``Other'', and provided an open-ended answer for the question. Since the sample size was small, and the answers contained similar concepts with multiple-choice answers, we did not find a theme in the open-ended answers. 
    
    When asked specifically what design aspect of ChatGPT participants liked compared to Google, responsiveness and interaction flow were the most common answers. The absence of ads was a popular choice too, especially among DUG, as it was the most chosen option for that group. It was the only option that didn't show a significant difference between the groups, meaning that both groups appreciate the ad-free experience of ChatGPT. Only 7 people chose ``None'' for this question (1 from DUB, 6 from DUG). Strikingly, when asked about what design aspect of Google they like compared to ChatGPT, 23 people answered ``None'' (18 from DUB, 5 from DUG). ChatGPT's design has a clear advantage, especially in the eyes of the participants who use both of them daily. Yet, the speed and responsiveness of Google were rated significantly higher by DUG. For design questions, we received 3 and 4 open-ended answers, respectively, for ChatGPT and Google preferences. Full answers to design and ChatGPT preference questions can be found in Appendix~\ref{supp_material}.
    
  \begin{figure*}[h]
        \centering
        \includegraphics[width=\linewidth]{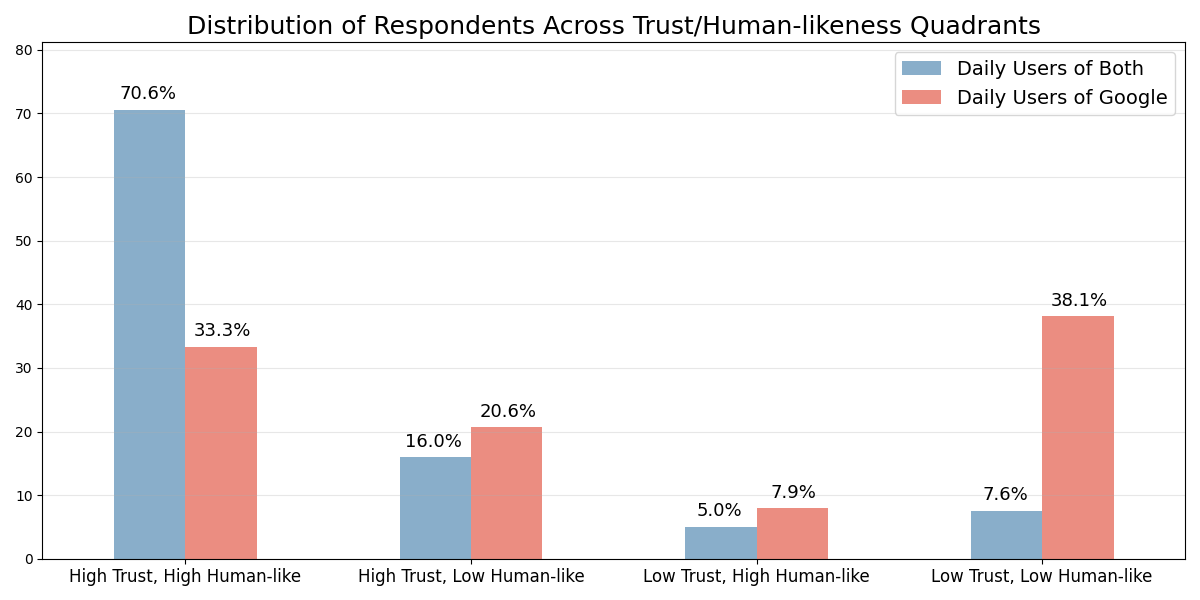}        \caption{Inner group splits based on trust and perceived human-likeness. Most of the DUB group has high trust and high perceived human-likeness for ChatGPT, while the DUG group shows a split between high trust/human-likeness vs. low trust/human-likeness.}
        \Description{The figure presents a quadrant-based visualization of participants' responses, divided by levels of trust and perceived human-likeness toward ChatGPT. The DUB group is concentrated in the high trust/high human-likeness quadrant, indicating a strong alignment between perceived anthropomorphism and trust. In contrast, the DUG group displays a more dispersed distribution, with notable portions in both the high trust–high human-likeness and low trust–low human-likeness quadrants, suggesting a more polarized perception of ChatGPT within this group.}
        \label{fig:quadrant}
    \end{figure*}

    To better understand design preferences, we looked into the relationship between trust, human-likeness, and design choices. Participants from both groups can be placed into a quadrant by classifying them as high/low trust x high/low perceived human-likeness. 70\% of the DUB group had both high trust and high perceived human-likeness for ChatGPT. Figure~\ref{fig:quadrant} shows an inner group split for DUG, where 38\% were low in both dimensions, but 33\% had high trust and high perceived human-likeness for ChatGPT.
    
    DUG sub-groups show a stark difference in their attitude towards ChatGPT; people who trust ChatGPT and perceive it as human-like value its interaction flow and interface simplicity at significantly higher rates than those who neither trust it nor find it human-like. For the low-trust/low human-likeness group, the only noticeable ChatGPT design feature they like is the absence of ads. Sub-groups also diverge in their willingness to trade off factuality. High-trust/ high human-likeness group is more willing to make the trade-offs, especially for more human-like and conversational interactions.

    \subsection{Age Groups}

    \begin{table}[]
      \centering
      \begin{tabular}{llc}
        \toprule
        \textbf{Attribute 1} & \textbf{Attribute 2} & \textbf{$p$‑value$^{\dagger}$} \\
        \midrule
        Ease‑of‑use     & Human‑likeness        & 0.005$^{*}$ \\
        Ease‑of‑use     & Conversationality     & 0.006$^{*}$ \\
        Human‑likeness  & Conversationality     & 0.664       \\
        \bottomrule
        \multicolumn{3}{l}{\footnotesize $^{\dagger}$Wilcoxon signed‑rank test (two‑tailed). $^{*}\,p<.05$.}
      \end{tabular}
      \caption{Paired comparisons of factuality trade‑offs for ChatGPT attributes among Daily Users of Google. The willingness to trade off factuality for ease-of-use is significantly different than conversationally and human-likeness.}
      \label{tab:DUG_pairs}
    \end{table}
    
    No statistically significant difference was found between genders for any question, while age groups show some significant differences, especially in usage patterns and trust. Adults over 55 use ChatGPT less frequently, especially compared to the 18-30 group. Despite their usage frequency, younger people are not the ones who trust ChatGPT the most. Adults between the ages of 30-40 have the highest trust for ChatGPT. They are also the ones who perceive ChatGPT to be human-like the most. Adults over 55 are the ones least willing to trade off factuality for any construct, while participants between 30-55 have a higher agreement with trade-offs, for all 3 constructs. Table~\ref{tab:age_group_means} shows the means of age groups for trust, human-likeness, and trade-off questions. 

    Extending the quadrant analysis to age groups, we did not find a significant difference between DUB sub-groups, as they all overwhelmingly fall into the high trust/high human-likeness group (see Figure~\ref{fig:quadrant}). Slight differences show a more positive perception in the middle age group (30-55). For DUG, the age group of 40-55 has an overwhelmingly high trust and human-likeness perception towards ChatGPT, as 61\% of the group falls into that quadrant. Interestingly, 18-30 DUG members fall more into the low/low quadrant. However, it should be noted that they don't have enough members (nine 18-30 olds in DUG) to provide meaningful insights.
    
\section{Discussion}

    Our results shed light on some interesting findings, particularly related to: a) identifying what kinds of users interact with both platforms, b) the relationship between their trust levels and perceived human-likeness, and c) what they like/expect about the user interfaces. 
    
    Regarding user groups, despite its popularity, ChatGPT has not replaced web search entirely: The DUB group shows that even the most frequent ChatGPT users also use Google on a daily basis. Still, this group seemed to form a personal relationship with ChatGPT, as they perceive it as human-like, and have high levels of trust. They also appreciate the interaction flow and the personalized interaction ChatGPT provides. DUB seems to be more pragmatist, as they are much more willing to sacrifice factuality for more personalization and ease-of-use. They also have trust in Google, even higher than DUG. 

    Personality (specifically, openness to technology) affects ChatGPT usage \cite{winds_of_change}. This is in line with our user group findings, as daily users of Google seem not to be ready to embrace using a conversational assistant like ChatGPT for information search, especially people who have low trust and low perceived human-likeness towards ChatGPT. Daily users of Google with high trust and high human-likeness have a more positive impression of ChatGPT, and appreciate its human-likeness and interaction flow despite not interacting with it daily. For participants with low trust and perceived human-likeness, personality may be a factor that affects their attitude toward ChatGPT.

    Regarding design preferences, the conversational nature of ChatGPT, backed up by personalized and human-like interactions, has a clear edge over traditional search engines. Both groups appreciate the interaction flow of ChatGPT, especially when their trust and perceived human-likeness levels are high. Extending this observation with our trade-off questions, we see that a smoother interaction might be worth trading off factuality for most people. The bond a user forms with the tool seems to play a role in their perception of the interface, as people with low trust and low human-likeness don't think interacting with ChatGPT has advantages over Google. 
    
    Our results show that the overtrust issue can get worse with the trust/human-likeness relationship people form with ChatGPT. However, in the studies that noticed this problem, participants were asked to interact with either a conversational interface or a traditional one. It means that they didn't have the opportunity to verify their responses using both tools. An emerging interaction phenomenon is to use a chatbot as a starting point for search, then verify the results with web search \cite{psychological_distance}. This interaction pattern seems to fit DUB. The willingness to trade off factuality for a better experience, especially in the DUB group, is related to the fact that people use traditional methods to fact-check conversational search results. Therefore, since users obtain factuality externally, they might value a human-like conversational experience more. 

    The lower trust and usage of ChatGPT among the 55+ group is in line with the previous literature, regarding the slower adoption of new tools in elderly people \cite{elderly_intention}. The difference with potentially higher implications seems to be between younger (18-30) and middle-aged adults (30-50). Young adults are the ones using ChatGPT the most, but their trust levels aren't the highest. 30-40 have the highest trust, and they agree the most that ChatGPT has human-like qualities. Younger adults might have a better understanding of hallucinations and the risk of ChatGPT generating incorrect answers due to their experience. Middle-aged adults, on the other hand, have higher trust despite lower usage frequencies, which may make them more vulnerable to overtrust issues.     

\section{Conclusion}

    In this paper, we conducted an exploratory study to understand how design, human-likeness, and trust affect people's perception of conversational search interfaces like ChatGPT, compared to traditional search engines. Our results uncover two main user groups by their usage frequencies, and we show how they differ in terms of trust and design preferences. The trust/human-likeness quadrants we discovered suggest that people appreciate the interaction and user interface more if they believe the tool is trustworthy and human-like. Furthermore, we explore the overtrust phenomenon and how it may be connected to the perceived human-likeness of conversational interfaces. Our findings show the importance of personalization and human-likeness for designing conversational interfaces to improve the interaction flow. For future work, our exploratory findings can form the background for predictive work. Even though we identify a relationship between trust, human-likeness, and design choices, the nature and the direction of the relationships are open questions. Furthermore, both trust and human-likeness have many dimensions, and future work should explore those dimensions in detail with multi-item questionnaires. Finally, our findings regarding user groups and trust/human-likeness quadrants would be better understood with more participants and more demographic variables like education level and profession.

\begin{acks} Anonymized
\end{acks}

\bibliographystyle{ACM-Reference-Format}
\bibliography{main}

\appendix

\section{Survey Questions}
\label{survey_questions}
\begin{enumerate}

\item How frequently do you use ChatGPT?
\begin{itemize}
    \item Multiple times a day (3+)
    \item A couple of times a day (1–2)
    \item Weekly
    \item Monthly
\end{itemize}

\item How frequently do you use Google?
\begin{itemize}
    \item Multiple times a day (3+)
    \item A couple of times a day (1–2)
    \item Weekly
    \item Monthly
\end{itemize}

\item I find Google to be trustworthy.
\begin{itemize}
    \item Strongly disagree
    \item Somewhat disagree
    \item Neither agree nor disagree
    \item Somewhat agree
    \item Strongly agree
\end{itemize}

\item I find ChatGPT to be trustworthy.
\begin{itemize}
    \item Strongly disagree
    \item Somewhat disagree
    \item Neither agree nor disagree
    \item Somewhat agree
    \item Strongly agree
\end{itemize}

\item Which platform do you trust more?
\begin{itemize}
    \item Google
    \item ChatGPT
\end{itemize}

\item ChatGPT has a personality or human-like qualities.
\begin{itemize}
    \item Strongly disagree
    \item Somewhat disagree
    \item Neither agree nor disagree
    \item Somewhat agree
    \item Strongly agree
\end{itemize}

\item What are the reasons you prefer ChatGPT over Google? (Select all that apply)
\begin{itemize}
    \item It feels more human-like
    \item It's easier to use and more intuitive
    \item I feel more engaged and understood
    \item It provides more personalized and tailored responses
    \item It's faster for certain types of queries
    \item It gives me clearer answers
    \item I trust it more for certain types of information
    \item None
    \item Other (please specify)
\end{itemize}

\item What design aspects of ChatGPT do you like better compared to Google? (Select all that apply)
\begin{itemize}
    \item Simplicity and cleanliness of the interface
    \item Speed and responsiveness
    \item Visual layout of results
    \item Presence or absence of ads
    \item Interaction flow when searching for information
    \item Mobile usability
    \item None
    \item Other (please specify)
\end{itemize}

\item What design aspects of Google do you like better compared to ChatGPT? (Select all that apply)
\begin{itemize}
    \item Simplicity and cleanliness of the interface
    \item Speed and responsiveness
    \item Visual layout of results
    \item Presence or absence of ads
    \item Interaction flow when searching for information
    \item Mobile usability
    \item None
    \item Other (please specify)
\end{itemize}

\item I would prefer ChatGPT over Google for information search because it is easier and more convenient to use, even if it sometimes provides incorrect or misleading information.
\begin{itemize}
    \item Strongly disagree
    \item Somewhat disagree
    \item Neither agree nor disagree
    \item Somewhat agree
    \item Strongly agree
\end{itemize}

\item I would prefer ChatGPT over Google for information search because it feels more human-like and personal, even if it sometimes provides incorrect or misleading information.
\begin{itemize}
    \item Strongly disagree
    \item Somewhat disagree
    \item Neither agree nor disagree
    \item Somewhat agree
    \item Strongly agree
\end{itemize}

\item I would prefer ChatGPT over Google for information search because it is conversational, even if it sometimes provides incorrect or misleading information.
\begin{itemize}
    \item Strongly disagree
    \item Somewhat disagree
    \item Neither agree nor disagree
    \item Somewhat agree
    \item Strongly agree
\end{itemize}
\end{enumerate}

\section{Supplemental Materials}

\label{supp_material}
    \begin{figure}[hpt]
        \centering
        \includegraphics[width=\linewidth]{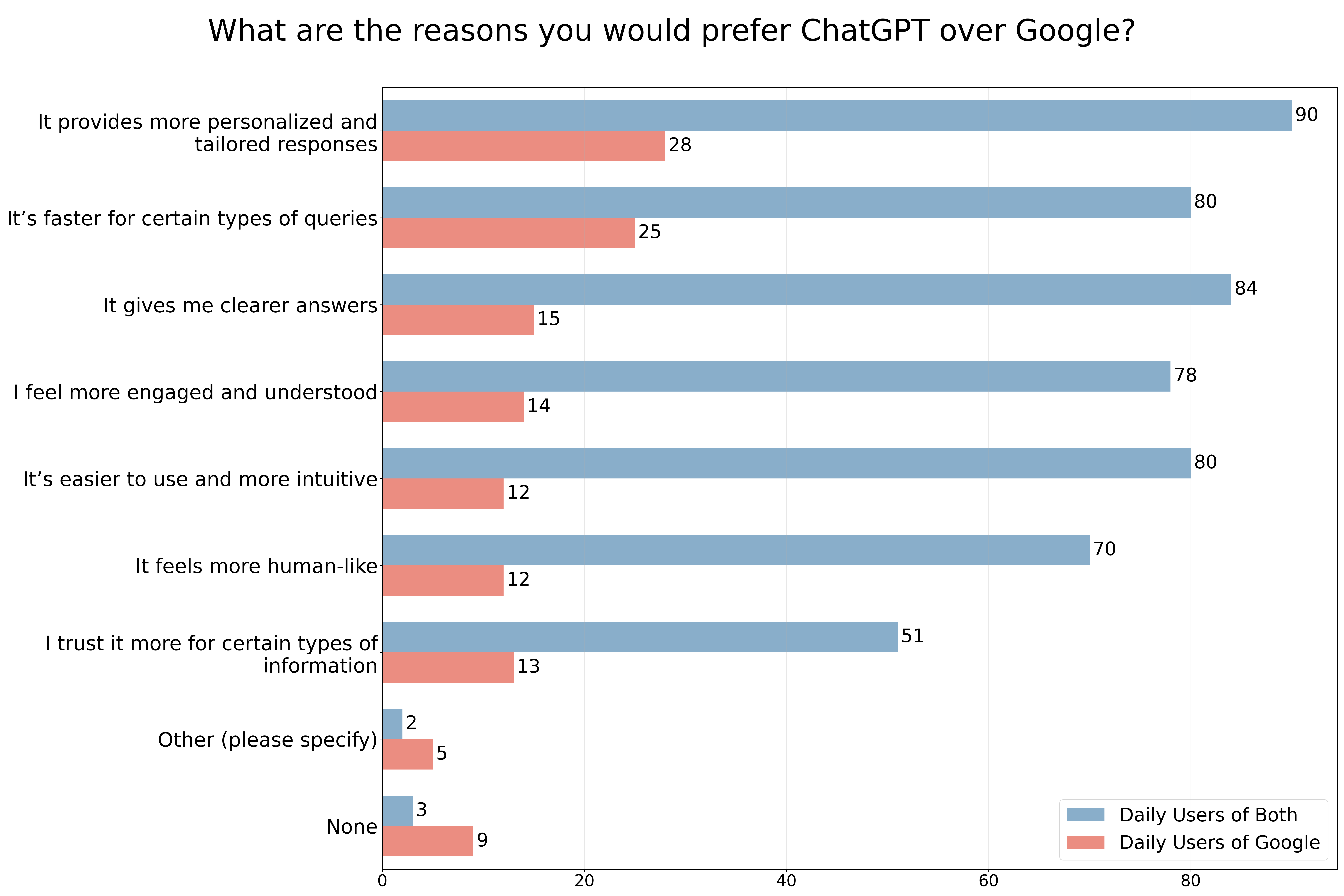}
        \caption{Reasons people prefer ChatGPT over Google.}
        \label{fig:chatgpt_preference}
        \Description{A horizontal bar chart titled 'What are the reasons you would prefer ChatGPT over Google?'. The chart compares responses from two groups: Daily Users of Both and Daily Users of Google. The most common reasons among Daily Users of Both include: 'It provides more personalized and tailored responses' (90), 'It gives me clearer answers' (84), 'It’s faster for certain types of queries' (80), 'It’s easier to use and more intuitive' (80), and 'I feel more engaged and understood' (78). Daily Users of Google showed significantly lower response counts across all reasons. Very few respondents selected 'None'.}
    \end{figure}
        
      \begin{figure}[hpt]
        \centering
        \begin{subfigure}{\linewidth}
            \centering
            \includegraphics[width=\linewidth]{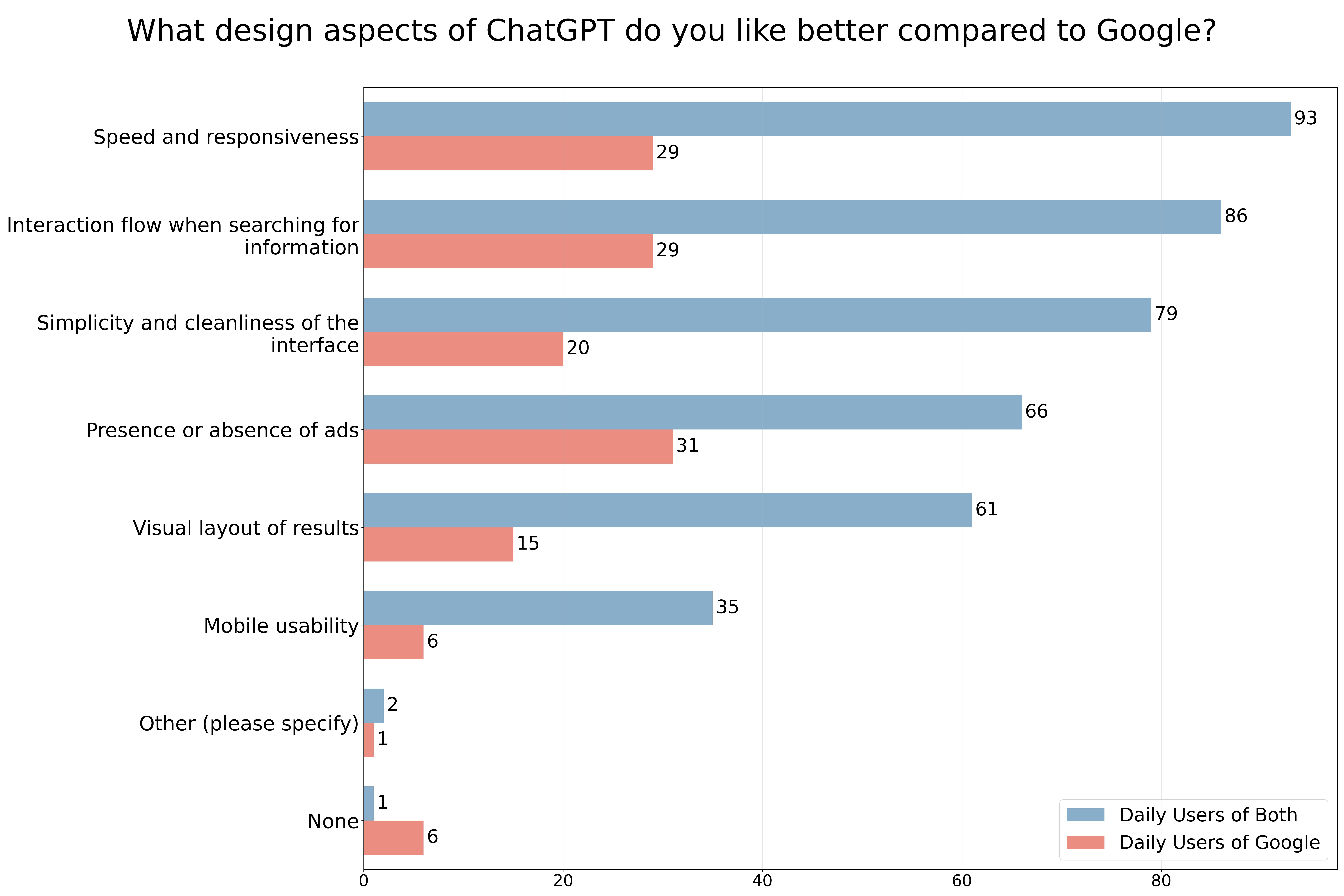}
            \caption{Design aspects of ChatGPT that are preferred over Google}
            \label{fig:design_ChatGPT}
        \end{subfigure}
            
        \begin{subfigure}{\linewidth}
            \centering
            \includegraphics[width=\linewidth]{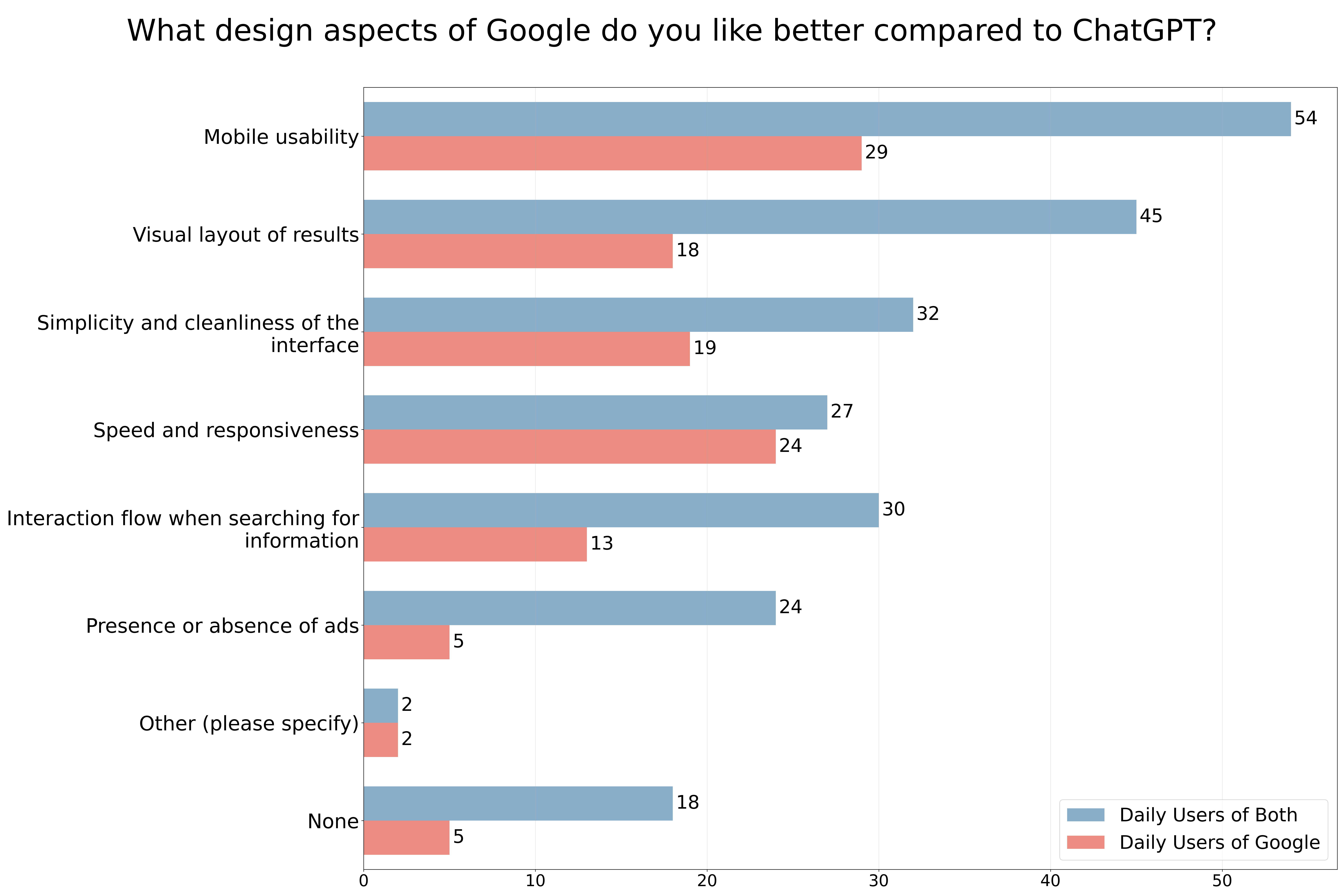}
            \caption{Design aspects of Google that are preferred over ChatGPT}
            \label{fig:design_Google}
        \end{subfigure}
        \caption{Results of multiple-choice questions regarding which design aspects of ChatGPT/Google are liked over the other one. Looking at the numbers, ChatGPT had a clear advantage regarding design.}
        \label{fig:design}
        \Description{Two horizontal bar charts displaying responses to multiple-choice questions about preferred design aspects of ChatGPT and Google. The first chart shows features of ChatGPT that users prefer over Google, with high scores for simplicity, speed, interaction flow, and absence of ads. The second chart shows features of Google that users prefer over ChatGPT, with lower overall selection rates, but notable mentions for speed and visual layout. Overall, ChatGPT is favored more strongly in terms of design.}
    \end{figure}
    
    \begin{table*}[]
      \centering
      \begin{tabular}{lcccc}
        \toprule
        \textbf{Item} 
          & \textbf{18–30} 
          & \textbf{30–40} 
          & \textbf{40–55} 
          & \textbf{55+} \\
        \midrule
        How frequently do you use Google?  & 3.619 & 3.667 & 3.673 & 3.647 \\
        How frequently do you use ChatGPT?   & 3.167 & 2.981 & 2.731 & 2.559 \\
        I find Google to be trustworthy.  
             & 4.143 & 4.259 & 3.942 & 4.000 \\
        I find ChatGPT to be trustworthy. 
             & 3.810 & 4.259 & 3.788 & 4.147 \\
        ChatGPT has a personality or human‑like qualities. 
             & 3.786 & 3.833 & 3.442 & 3.412 \\
        I would prefer ChatGPT over Google because it is easier to use, \\
        despite incorrect or misleading information.  
             & 3.357 & 3.111 & 3.096 & 2.765 \\
        I would prefer ChatGPT over Google because it is humanlike, \\
        despite incorrect or misleading information.  
             & 3.262 & 3.222 & 3.269 & 2.912 \\
        I would prefer ChatGPT over Google because it is conversational, \\
        despite incorrect or misleading information.  
             & 2.976 & 3.352 & 3.250 & 2.853 \\
        \bottomrule
      \end{tabular}
      \caption{Mean survey scores by age group for trust, human‑likeness, and trade‑off questions. For the usage frequency questions, answers are labeled from 4 (Multiple times a day) to 1 (Monthly).}
      \label{tab:age_group_means}
    \end{table*}

\end{document}